# Elevated temperature effects (T > 100 °C) on the interfacial water and microstructure swelling of Na-montmorillonite

Wei Qiang Chen*, Majid Sedighi*, Florent Curvalle, and Andrey P Jivkov*

School of Engineering, The University of Manchester, Manchester, M13 9PL, United Kingdom

*Corresponding authors:

Weiqiang.Chen@manchester.ac.uk;

Majid.Sedighi@manchester.ac.uk;

Andrey.Jivkov@manchester.ac.uk

**Abstract**

Montmorillonite-based barriers are key elements of the engineered barrier systems (EBS) in geological disposal facilities (GDF). Their performance at temperatures above 100 °C is not sufficiently understood to assess the possibility of raising the temperature limits in GDF designs that could reduce construction costs and $CO_2$ footprint. The present work provides new fundamental insights through molecular dynamics (MD) simulations of Na-montmorillonite's water-clay interactions and swelling pressure at temperatures 298–500 K and basal spacings of 1.5–3.5 nm. At temperatures above 100 °C, the swelling behaviour is governed by the attractive van der Waals force and the repulsive hydration force instead of the repulsive electrostatic (double layer) force. The swelling pressure reduction with increasing temperature is related to the weakened hydration repulsion and electric double layer repulsion, which result from the deterioration of the interlayer water layer structure and the shrinkage of the electric double layer. The applicability and breakdown of the classic Derjaguin-Landau-Verwey-Overbeek (DLVO) theory at elevated temperatures are examined. By excluding the osmotic contribution in the DLVO theory, the summation of the van der Waals interaction in DLVO and an additional non-DLVO hydration interaction can predict our MD system's swelling under high temperatures. The findings of this study provide a fundamental understanding of the swelling behaviour and the underlying molecular-level mechanisms of the clay microstructure under extreme conditions.

**Keywords**

1. Introduction

In deep geological disposal of nuclear waste, clays rich in montmorillonite mineral, commercially known as bentonite, are considered for buffer and backfill materials between canisters containing heat-generating high-level nuclear waste (HLW) and host rock. A key property of clays in geological disposal is their ability to maintain a high swelling pressure developed during re-saturation under the constrained conditions of the repository. To ensure the buffer/backfill fitness-for-purpose, this swelling pressure must be maintained under elevated temperatures for a long period of time. In current conceptual designs of disposal sites, the spacing between HLW canisters is constrained by the requirement that the temperature of the clay buffer/backfill is below 100 °C. Therefore, the research in the last 30 years on the hydration and swelling pressure development of clays has been focused on temperatures below 100 °C. There is a growing interest in the possibility of increasing this limit, which will allow for reducing the spacing between canisters, significantly lowering the cost and carbon footprint of geological disposal facilities' construction. However, very few studies on the clay behaviour at temperatures above 100 °C have been conducted. The current understanding from these is insufficient to make informed decisions by designers and regulatory bodies.

Real-time microscopic experiments that reveal the swelling process and capture the microstructural evolution of clay at elevated temperatures (T >100 °C) are not available. This is mainly due to the constraints on temperature ranges and complex experimental conditions at temperatures above boiling point [1]. It has been shown that samples of bentonite exposed to temperatures above 100 °C may experience significant mineralogical transformations such as illitisation and cementation [2, 3]. Depending on the density at saturation, exchangeable cation types, saline environment and other experimental conditions, increasing as well as decreasing swelling pressure at temperatures above 100 °C have been reported, e.g. [4]. The underlying microscopic mechanisms remain unclear. Different conjectures have related temperature effects on swelling pressure to structural and dynamical modifications of interlayer liquid, e.g., thermally increased diffuse layer thickness and thermally decreased dielectric constant and surface potential in pore fluid [5], thermally enhanced double layer repulsion [6], and thermally induced transport of interlayer water to the macropores [7].

The existing microscopic numerical studies on temperature effects have been mainly performed by molecular dynamics (MD) simulations and molecular-level grand canonical Monte Carlo (GCMC), e.g., [1, 8-13]. These have been focused on the microstructure swelling of Na-montmorillonite at temperatures below 100 °C (or 373.15 K), but some have also considered temperatures above 100 °C, reporting increasing as well as decreasing swelling pressure at such temperatures, e.g., [1, 8, 9, 11, 12]. These behaviours have been explained by enhanced thermal motion of interfacial water molecules [11], weakened hydration interactions due to destroyed interlayer water structure and reduced double-layer repulsions caused by enhanced ion correlation effects [12]. Temperature effects on the swelling pressure have also been found to be stronger for smaller basal spacings [8, 11]. The swelling results have been found to be consistent with the clay hydration enthalpies and the interlayer liquid structure [9]. These studies have advanced the understanding and quantification of the temperature effects on swelling and microstructural evolution of montmorillonite clay, such as the hydration state/energy/force of clay surfaces and interlayer cations, but a comprehensive microscopic physio-chemical explanation of the thermal effect is still lacking. The temperature effects under complex conditions encountered at practical engineering applications, such as those related to the presence of gases like water vapour, methane and carbon dioxide in the interlayer region of swelling montmorillonite clays [14], require further numerical investigations.

Presented in this paper is a molecular dynamics (MD) investigation of the swelling behaviour and the underlying microscopic mechanisms of a water-saturated Na-montmorillonite (Na-MMT, the major functional component of bentonite clays [1]) at elevated temperatures (>100 °C). One important contribution is the design of the MD system, where the boundary conditions capture the micro-macro water and chemical exchanges. To the best of our knowledge, this has been largely missed in previous studies, limiting their physical realism. A second contribution is the quantification of the swelling pressure development with temperature and basal spacing and the corresponding microscopic evolutions of the interlayer liquid structure, including the clay hydration/dehydration and the counterion solvation/de-solvation. A third contribution is the analysis of the applicability of the classical Derjaguin-Landau-Verwey-Overbeek (DLVO) theory at elevated temperatures, demonstrating when it breaks down, identifying the causes for the breakdown, and improving its prediction capability by further considering the hydration interaction.

## 2. Methods

**2.1 Molecular dynamics system**

The model system is shown in **Figure 1**. It consists of a water reservoir with two parallel saturated clay substrates in the centre. Two implicit walls interacting only with $Na^+$ ions are placed at the edges of the clay substrates to prevent the leakage of the counterions and to favour osmotic swelling instead of crystalline one [12]. The effects of these implicit walls are quantified in the **Supporting Information** and are found to be in favour of statistical stability [1, 10, 12]. To control the water reservoir pressure, two bounding pistons are added. The left piston is constrained, while the right one can move in the $x$ direction to apply the desired pressure. Periodic boundaries are applied in the three spatial directions, and some vacuum space is created at the end of the right piston to prevent periodicity conflict in the $x$ direction.

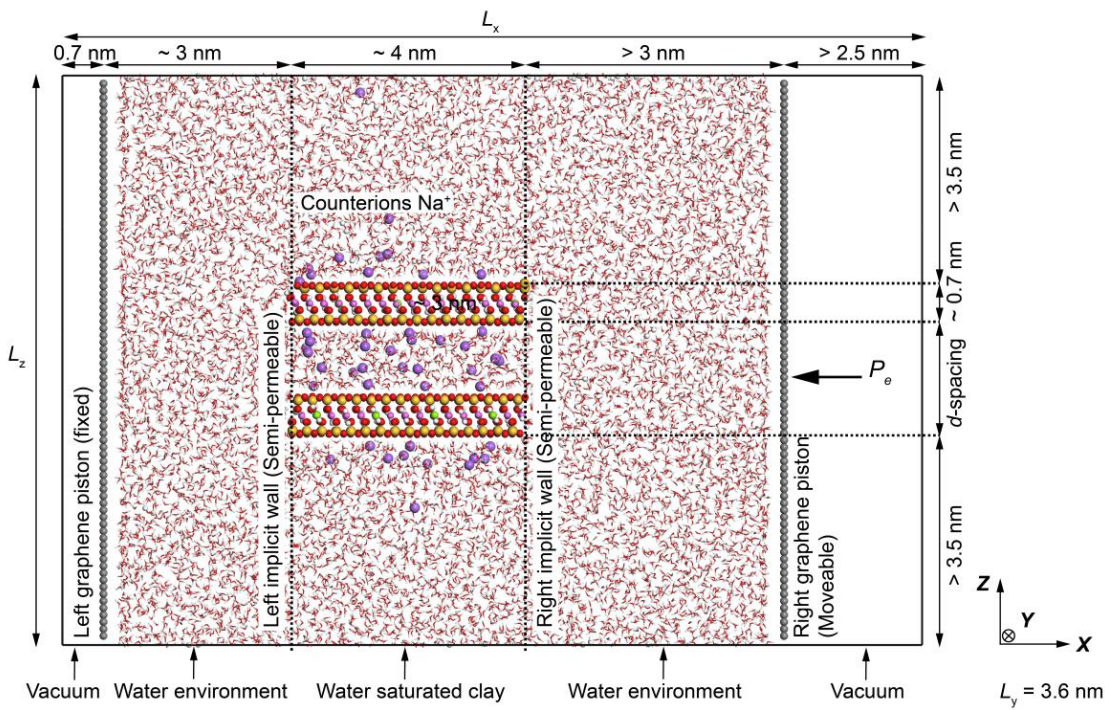

**Figure 1.** Schematic of the molecular dynamics system.

The dimensions of the system are nearly 15, 3.6, and 10 nm in the $x$, $y$, and $z$ directions, respectively. The unit cell of a Wyoming-type Na-montmorillonite with a chemical formula of $Na_{0.75}[Si_{7.75}Al_{0.25}][Al_{3.5}Mg_{0.5}]O_{20}(OH)_4$ was adopted to construct the clay substrates [12]. Both montmorillonite substrates are aluminosilicate with one octahedral layer wrapped by two tetrahedral layers. The supercell consists of a 64-unit cell with the isomorphic substitution of $Mg^{2+}$ for $Al^{3+}$ in the octahedral layer, and $Al^{3+}$ for $Si^{4+}$ in the tetrahedral layers following Loewenstein's rule [15]. This isomorphic substitution was uniformly distributed according to the model used in the Ngouana and Kalinichev [16]. Therefore, it results in a net negative charge, which is neutralised by the $Na^+$ ions in the interlayer space between the two implicit walls. The simulation domain contains 38900 atoms, including 11244 water molecules and 48 $Na^+$, represented by the purple spheres in **Figure 1**.

The swelling behaviour in the direction normal to the basal surface of MMT was investigated. The bottom clay substrate is fixed in the three spatial directions during the simulations, while the top one is tethered with a spring in the $z$ direction and constrained in the $x$ and $y$ directions. The spring model is depicted in **Figure 2**(a), where the spring follows Hooke's law with a spring constant $k$. The swelling pressure is directly linked to the $z$ displacement of the top clay substrate relative to its initial position. A displacement in the positive (negative) $z$ direction of the top clay substrate from its initial position represents expansion (shrinkage). The spring restoring force increases to restrain the swelling (contraction). The spring force is proportional to the variation of the d-spacing. Simulations were performed with spring constants in the range from 10 to 150 kJ/mol/nm$^2$ and with initial d-spacings in the range from 1.0 to 3.0 nm. The swelling pressure, $P_s$, was calculated as follows:

$$P_s = \frac{F_{swelling}}{S} = \frac{F_{spring}}{S} = \frac{k \Delta z}{S} \qquad (1)$$

where, $F_{swelling}$ is the swelling force, $S$ the surface area of the clay substrate, $\Delta z$ is the average variation of d-spacing from its initial value.

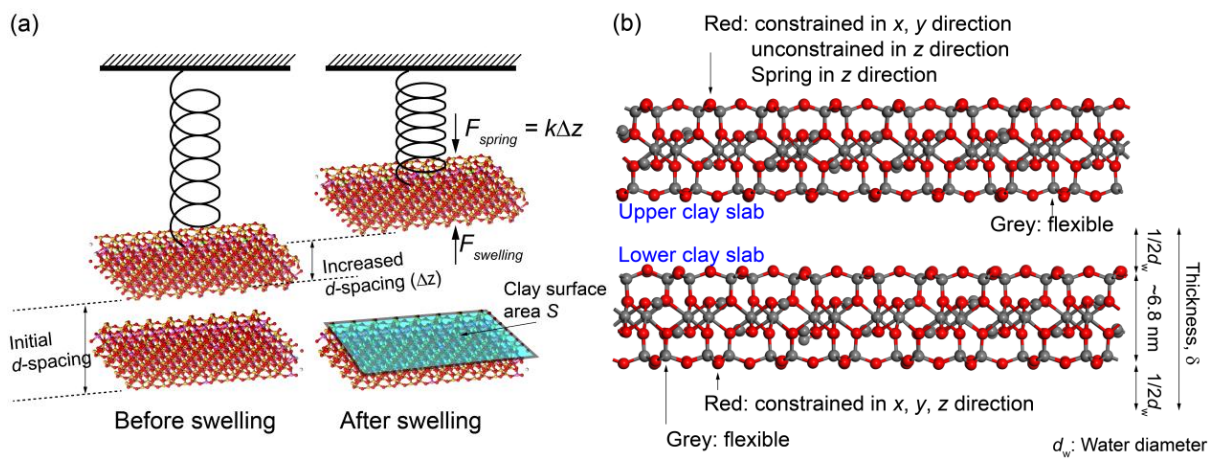

**Figure 2.** (a) The vertical spring is attached to the clay substrate to measure the swelling pressure, where the bottom clay substrate is constrained while vertical movement is allowed for the top one. (b) Specific sites on clay layers where the spring is attached.

Unlike previous studies where fully rigid clay layers [12] have been considered, the clay slabs in this model are partially rigid and partially flexible. **Figure 2**(b) shows that one part (bridging oxygens) was kept rigid while the other part was flexible. This system is more realistic because the fully rigid clay layers are completely incompressible, and they do not consider the thermal vibration of clay atoms, indicating a temperature of 0 K. However, additional calculations showed negligible differences in simulation results between these two setups, possibly because the flexible atoms (grey) reside inside the framework made up of the rigid atoms (red). Therefore, their effects on the interlayer liquid and swelling are negligible.

The use of different initial d-spacings and spring constants allows for analysing an adequate number of cases so that the final equilibrium swelling pressure is compensated by the spring force induced by different changes in the d-spacing. First, five different initial d-spacings (i.e., 1.0, 1.5, 2.0, 2.5, 3.0 nm) and three different spring constants (i.e., 10, 50, 150 kJ/mol/nm$^2$) were used to establish 5×3=15 initial MD systems. Second, each system was allowed to equilibrate, and the initial d-spacing would increase or decrease until the equilibrium d-spacing was attained, where the spring force equalled the swelling force. Third, the equilibrium d-spacings and corresponding swelling pressures from 15 cases were recorded to produce the relation between the swelling pressure and the basal d-spacing. For the same initial d-spacing, varying spring constants would produce different data points on the profiles of swelling pressure and d-spacing, as shown in **Figure 6**. In addition, a lower spring constant would cause a larger fluctuation of swelling pressure and d-spacing around their equilibrium values.

After deriving the relations between the swelling pressure and final equilibrium d-spacing, the corresponding liquid structure inside and outside the interlays will be analysed by fixing the upper and lower Na-MMT slabs at the equilibrium d-spacing. The derived relations between swelling pressure and the basal spacing should be independent of these two variables.

## 2.2 Simulation details

All MD simulations were run using LAMMPS [17]. The results are presented by the VMD [18] package. The ClayFF [19] force field was used to describe the system, and the atomic partial charges were re-assigned according to the previous study of Ngouana and Kalinichev [16]. The nomenclature for each atom type and the corresponding partial charges are shown in **Figure 3**. The rigid simple point charge (SPC) model was used to describe water molecules and implemented by the RATTLE [20] algorithm. The Lorentz-Berthelot combining rule was used to compute the Lennard-Jones (LJ) 12-6 parameters for unlike species. The cut-off distance of 1.2 nm was adopted for the non-bond interactions, and the particle-particle particle-mesh (PPPM) method was applied for the long-range Coulomb interactions with a precision of $10^{-5}$. These parameters have been widely proven to accurately describe the structural and dynamic properties of MMT/aqueous solutions interfaces [1, 10, 12, 21]. The pistons used are in graphene with the LJ 12-6 parameters taken from a previous study by Chen et al. [22].

First, a relaxation stage of 20 ns was computed in the $NVT$ ensemble at the specified temperature and environmental pressure ($T = 298/350/400/450/500K$ and $P_e = 50MPa$) to obtain the equilibrium state. These environmental factors, i.e., high temperature and high pressure, are consistent with the extreme conditions experienced by the clay barrier [23]. Second, a simulation run of the production stage of 30 ns was carried out on the same ensemble for data analysis. The system temperature was controlled via the Nose-Hoover thermostat [24] with a temperature damping parameter of 0.2 ps. Newton's equations of motion are computed by the Velocity Verlet algorithm [25] with a timestep of 2 fs and an interval of 200 fs for statistical data collection. The atomic trajectories were recorded in an interval of 8 ps. The interactions between the sodium ions and implicit walls were computed with the harmonic potential using a spring constant of 5000 kJ/mol/nm$^2$ and a cut-off distance of 0.25 nm. The radial distribution function (RDF) and the coordination number (CN) were computed for specific atom pairs, and the potential of mean force (PMF) was determined according to $W(r) = -k_B T \ln(g(r))$ [26], where $g(r)$ is the radial distribution function. It is noted that the water in this study was always as liquid phase according to the pressure-temperature phase diagram of water.

Note, that the equilibration process for such a system to reach a steady state can be slow. To ensure the equilibration of our MD system was appropriately achieved, the equilibrium time for each simulation case was set to 20 ns, followed by 30 ns production time, as mentioned above. Previous studies [10, 12, 27] have shown that such an equilibrium time is sufficient, particularly under high temperatures where interlayer water molecules and counterions have higher mobility [28]. Moreover, each case was repeated three times with different initial configurations, and statistically, the same result was obtained. Finally, several cases were repeated with longer simulation times up to 150 ns, and the result obtained showed negligible difference from those with 30 ns.

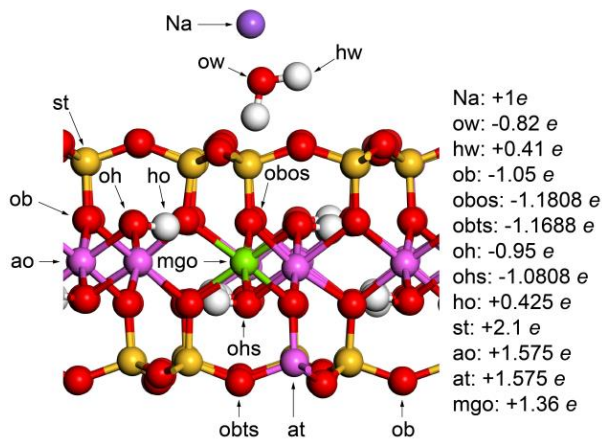

**Figure 3.** The definition for the atom types and their partial charges in the Na-MMT.

## 3. Results and discussion

### 3.1 Impacts on the interfacial liquid

This section only presents results for the basal spacing of 1.95 nm, and the corresponding results for the basal spacings of 1.53 nm and 2.2 nm can be found in **Figures S1–S4** in the **Supporting Information**.

**Interlayer counterions and water molecules:** The liquid structure inside and outside the interlayer space of the Na-MMT nanochannels at different temperatures has been quantified. **Figure 4**a–b shows the distribution profiles for the number density of the sodium counterions and water molecules.

First, it was found that there is a deterioration, suppression, and destruction of the layering structure of interfacial water with increasing temperature, which can be observed by the flattened peaks and valleys of the profiles under higher temperatures. The three peaks at 298 K in the water density profiles in **Figure 4**(a–b) indicate the presence of precisely three water layers (3WL). The reduction of these peaks with increasing temperature shows that the interlayer space requires fewer water molecules and that a certain number of water molecules have been squeezed out of the interlayer space due to thermal expansion. The hydration degree and hydration force of the Na-MMT is reduced with rising temperature. **Figure 4**(a–b) and **Figure S1–S2**(a–b) show more peaks in the water density profiles for larger basal spacings, and their number increases with increasing temperature. It can be observed in **Figure 4**(a–b) that there is a transition from three water layers (3WL) at ambient temperature to four water layers (4WL) at elevated temperature, because the hydration films absorbed on the two clay surfaces become thinner, and more peaks and valleys with flattened oscillations appear in the same size of the pore. The reason is that at ambient temperature, interlayer water molecules form distinct, stable, and ordered adsorbed water layer structures at the clay surface. At high temperatures a certain amount of water molecules has been squeezed out of the interlayer space which weakens the hydration effect of the clay surface. Fewer water molecules directly adsorb on the clay surface as the first water layer, reflected by the lowered first peak value of the interlayer water density profile in **Figure 4**(a–b). In this case, the original second water layer can penetrate the first water layer and get closer to the clay surface to occupy the energetically favourable sites, with the same case for other water layers. A higher entropy of interlayer water molecules at high temperatures can be observed in **Figure S2**(b), where the water density profile at 500 K in the middle of the nanochannel becomes flat, i.e., the water layer structure becomes weak and a bulk water environment is formed. This temperature-induced transition of the hydration state of Na-montmorillonite has also been observed experimentally in a previous study [29].

Second, it has been pointed out in a previous study [30] that at ambient temperature, the basal spacings for Na-montmorillonite are in the range of 1.15–1.25 nm for the monohydrated (1WL), 1.45–1.55 nm for the bi-hydrated (2WL), 1.8–1.9 nm for tri-hydrated clay (3WL) and less or equal to 2.2nm for tetra-hydrated clay (4WL). The results presented here are consistent with these observations. The studied d-spacing and the corresponding number of water layers are further shown in **Figure 6**(c).

Third, the shrinkage and weakening of the electric double layer with increasing temperature can be observed by the weakening oscillations, i.e., flattened peaks/valleys of $Na^+$ number density profiles inside the nanochannel under higher temperatures. As shown in **Figure 4**(b), the counterions form inner-sphere surface complexes (ISSC), outer-sphere surface complexes (OSSC), and diffuse swarm (DS) species [31] in the interlayer region, where counterions are directly bound to the clay surfaces in ISSC, while they retain intact solvation shell in OSSC and DS. **Figure 4**(b) shows that the central peak in the middle of the channel, as well as their neighbouring peaks of the $Na^+$ density profile decrease with increasing temperature, which shows that the counterions in OSSC and DS tend to dehydrate themselves, get closer to the clay surfaces and form ISSC. Therefore, under high temperatures, the surface charge will be more effectively screened by the approaching counterions, thus reducing the thickness of the electric double layer.

**Electrical properties of the interlayer space: Figure 4**(c) shows the charge density distribution profiles, $\rho_q(z)$, of the interlayer liquid and external liquid under different temperatures for the basal spacing of 1.95 nm, where the liquid reservoir is excluded. A weakened electrical double layer with rising temperature can be observed from the reduced oscillation amplitudes of $\rho_q(z)$. Integrating twice Poisson's equation $\frac{d^2\phi(z)}{dz^2} = -\frac{\rho_q(z)}{\varepsilon_0}$ [32], gives the electrical potential $\phi(z)$, as shown in **Figure 4**(d), where the contributions from the whole system, i.e., ion, water, and clay, and the periodic boundary conditions are considered. The results show that the magnitudes of the calculated electrical potentials $\phi(z)$ decrease with increasing temperature, indicating weakening of the electric double layer. This finding is consistent with the experimental observation of a previous study [33], where the surface potential of bentonite particles dispersed in the water and heated at different temperatures are measured. See **Figure S1–S2** for the basal spacings of 1.53 nm and 2.2 nm, which produce exactly the same conclusions.

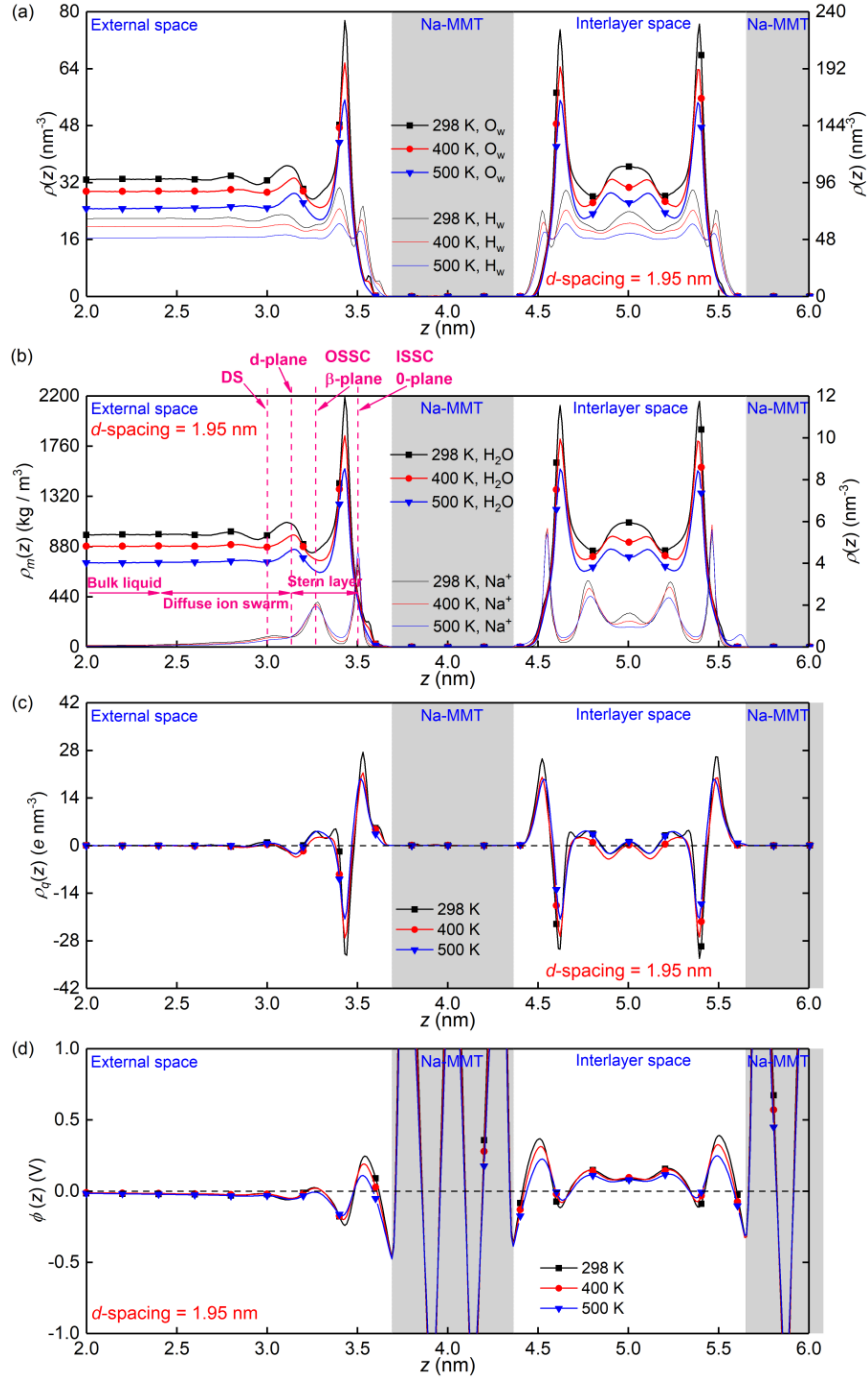

**Figure 4.** (a–b) The number/mass density distributions of water and counterions (Na$^+$) inside and outside the interlayer space of montmorillonite with a d-spacing of 1.95 nm at 50 MPa and different temperatures, where inner-sphere surface complexes (ISSC), outer-sphere surface complexes (OSSC), and diffuse swarm (DS) species and the corresponding adsorption planes of 0-, β-, d-plane have been indicated. (c) The corresponding charge density profiles. (d) The corresponding electrostatic potential of the system along the $z$ direction.

**Solvation behaviour of the interlayer counterions: Figure 5** shows the calculated radial distribution function (RDF) and coordination number (CN) of interlayer cations with water oxygens and Na-MMT oxygens for the basal spacing of 1.95 nm (see **Figure 3** for the nomenclature of different oxygen types). The potential of mean force (PMF) is also computed for the Na − ow pairs.

The counterions cannot be considered as simple point charges due to the hydration action of the surrounding water molecules [34]. The size of dehydrated ions is smaller, allowing them to approach the charged surfaces easily and closely. **Figure 5**(a) shows two peaks on the RDFs of Na − ow pairs, which represent two hydration layers surrounding the cations, and the peak heights show the number of water molecules in the corresponding hydration layers. The first peaks decrease with increasing temperature, indicating dehydration of counterions. **Figure 5**(b)–(d) show the amount of adsorbed counterions on clay surfaces. This amount increases with increasing temperature. The conclusion is that with increasing temperature, an increasing number of sodium ions are absorbed on the montmorillonite surfaces instead of coordinating with water molecules due to the dehydration of counterions. This is consistent with the results in the previous subsection that sodium ions move closer to the surfaces with temperature increasing. In summary, sodium ions ($Na^+$) tend to dehydrate and come closer to the charged clay surfaces by reducing the number of water molecules they are coordinated with and by coordinating instead with the clay surface. **Figure 5**(e) further shows that the well depths of the PMF profiles increase with increasing temperature, which indicates reduced hydration ability of the sodium ions with rising temperature.

See **Figure S3–S4** for the basal spacings of 1.53 nm and 2.2 nm, which produce exactly the same conclusions.

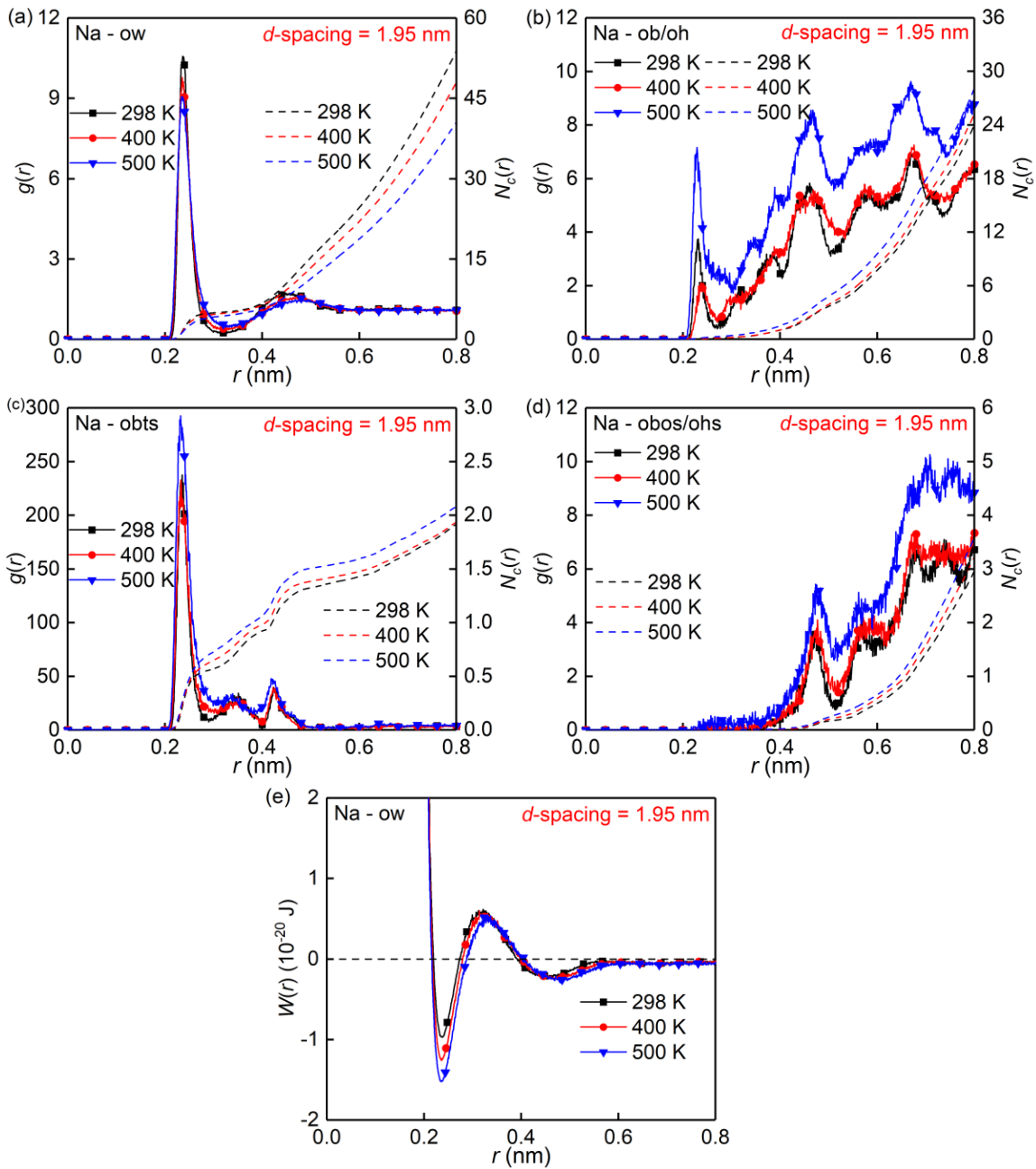

**Figure 5.** (a–d) The radial distribution functions (RDF, solid lines) and coordination numbers (CN, dashed lines) of different pairs with d-spacing of 1.95 nm at 50 MPa and different temperatures. (e) The potential of mean force (PMF) for the Na − ow pair.

### 3.2 Impact on microstructure swelling

The scattered points in **Figure 6**(a–c) show the variations of the calculated swelling pressure ($P_s$) with the equilibrium basal spacing ($d$) of Na-MMT under different temperatures (and environment pressure of 50 MPa), as well as their empirical fitting in **Figure 6**(a), theoretical prediction by the classic DLVO theory in **Figure 6**(b), and theoretical prediction by the extended DLVO theory further considering the hydration interaction in **Figure 6**(c). Four calculation examples of swelling pressure and basal spacing at equilibrium states can be found in **Figure S5** in the **Supporting Information**, together with four corresponding simulation cases without implicit walls, which are only used to demonstrate that the setup with implicit walls can achieve better statistical stability. The error bars were calculated by repeating each simulation case three times with different initial configurations and obtaining the standard deviation. The swelling pressure is of the same order of magnitude as experimentally measured values, e.g., refs. [4, 5, 35]. A direct comparison is not feasible because of the differences between the nanoscale simulations and the macroscopic experiments, such as the size of clay platelets, the number of stacked clay platelets, the multiscale porosity and saturation state, the volume fraction of the non-swelling minerals in the composition, the stacking and orientation pattern of clay particles. However, the MD simulations reveal the microscopic mechanisms responsible for the observed trends.

**Empirical fitting: Figure 6**(a) shows that with increasing temperature and the same basal spacing, the swelling pressure of Na-MMT drops. The behaviour generally changes from expansion to shrinkage when the temperature increases from 298 K to 500 K. The calculated thermally induced monotonical decrease of the swelling pressure is in agreement with previous experimental studies [4, 35-37]. Pusch has attributed it to the existence of less stable interlayer water at a higher temperature [36], which is consistent with the molecular-level quantifications in the previous section. A previous study [33] also found the enhanced agglomeration of montmorillonite particles dispersed in water under high temperatures, which agrees well with our results above that the repulsion between clay particles weakened as temperature increased.

It is noted that the shape of the profiles undergoes a transition from the repulsive-force-dominant shape (predominantly electrostatic double layer interactions in nature) to the attractive-force-dominant shape (predominantly Van der Waals interactions in nature). This is shown by the dashed lines of **Figure 6**(a), which are plots of an exponential function, $P_s = k_1 e^{-k_2 d} + k_3$, fitted to the data from 298–450 K cases, and of a Lennard-Jones 12-6 potential function, $P_s = 48k_1(\frac{k_2^{12}}{d^{13}} - 0.5\frac{k_2^6}{d^7})$, fitted to the data of 500 K cases ($k_1$, $k_2$, and $k_3$ are the fitting parameters). This change from repulsive to attractive behaviour can be explained as follows. As the temperature increases, the surface charge is gradually screened by the counterions, and the structural thickness of the diffuse double layer decreases (see the previous section). This leads to a gradual replacement of the electrostatic double layer interactions by the van der Waals interactions. Similar explanations have been used to discuss the ion concentration effect on swelling pressure, e.g., ref. [35], where increasing the ionic strength of the pore fluid results in a decrease in the swelling pressure due to a decrease in the thickness of the electric double layers and their overlapping. This has been proven mechanistically in the previous section.

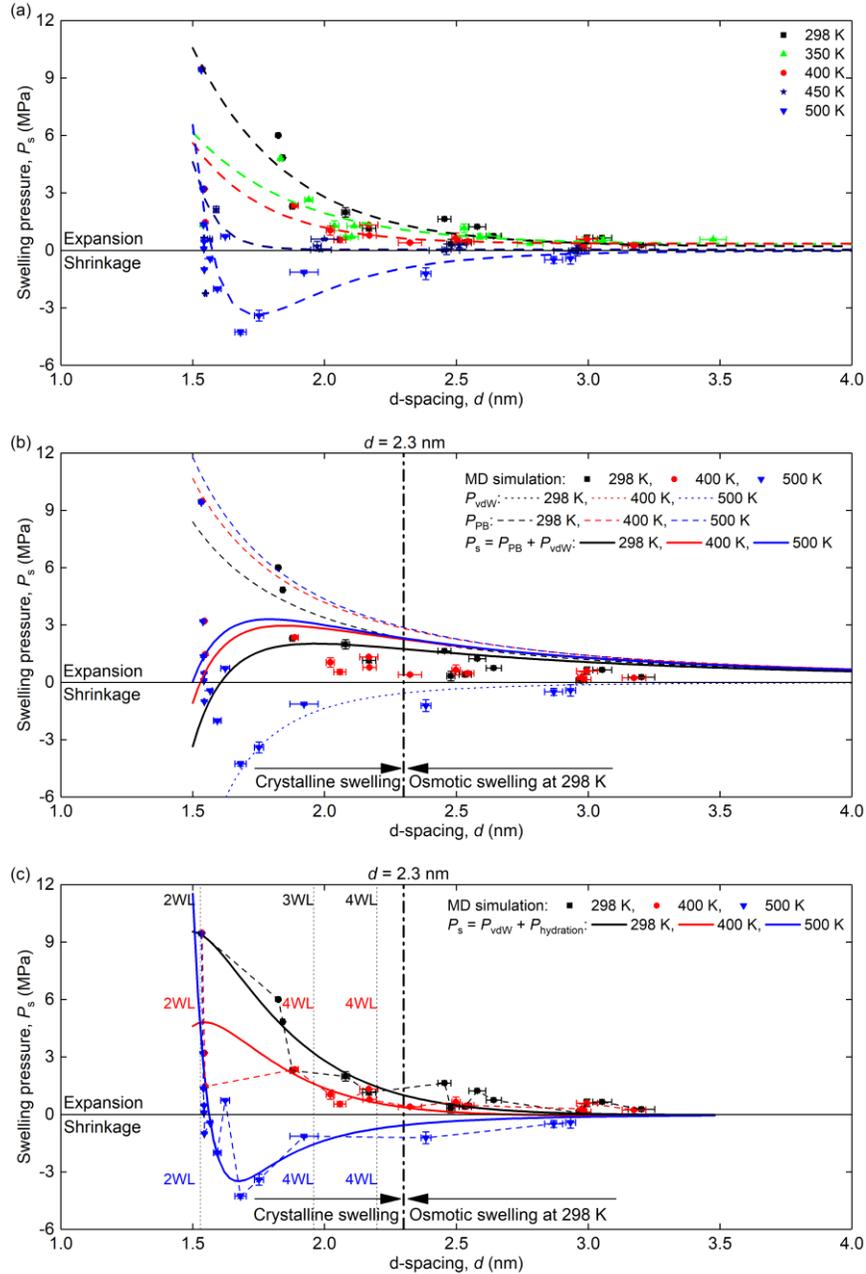

**Figure 6.** The swelling pressure at an environment pressure of 50 MPa and different basal spacings and temperatures. (a) The raw data and the fitted trend lines using empirical equations. (b) The comparison of swelling pressure between MD simulations and DLVO theory. (c) The oscillating behaviour of the raw data, the theoretical prediction of the extended DLVO model considering the hydration interaction, and the specific d-spacings with the corresponding water layer numbers indicated. The vertical dash lines show the basal spacing studied and analyzed in the previous section, and the black dash-dotted line indicates the basal spacing $d = 2.3$ nm separating the crystalline swelling and the osmotic swelling.

**DLVO model predictions:** The applicability and breakdown of the classic Derjaguin-Landau-Verwey-Overbeek (DLVO) theory [38, 39] at elevated temperatures are examined further, and the causes of its breakdown are identified hereafter. **Eqns. (S1)–(S12)** [30, 32, 40, 41] in the **Supporting Information** present the classic DLVO theory predicting the disjoining pressure between two charged planar surfaces, confining a solution with only water and counterions. Using these equations, the distribution profiles of counterion number density and of electrostatic potential are presented in **Figure 7**(a) and (b), respectively. The corresponding parameters and their sources [12, 42] are shown in **Table S1** in the **Supporting Information**. **Figure 7**(a) shows that increasing the temperature leads to a reduction of counterion concentration in the middle of the slit nanochannel and to an increase of counterion concentration near the solid surfaces. These are consistent with the MD results presented in **Figure 4**(b). However, the electrical double layer structure shown in **Figure 4**(a–c) cannot be fully reproduced by **Figure 7**(a) because the specific interactions between water molecules, surface sites, and counterions, e.g., surface complexes and ion pairing [43], are not considered in the classic DLVO theory. **Figure 7**(b) shows that increasing the temperature leads to a decrease in the electrostatic potential near the solid surface. Note that the boundary condition represented by **Eq. (S4)** might not apply, i.e., $\phi(0) \neq 0$, when the basal spacing is narrow, therefore the electrostatic potential presented in **Figure 7**(b) is simply based on the assumption of $\phi(0) = 0$.

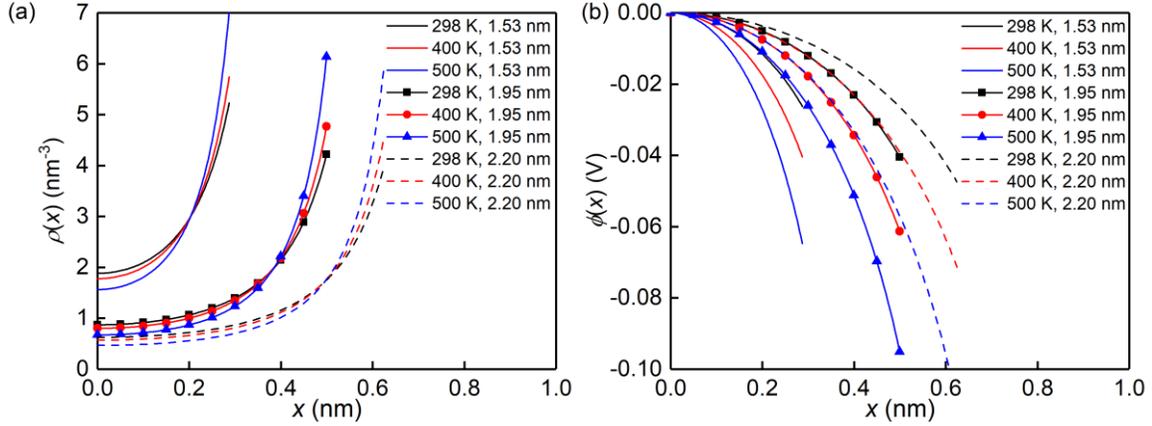

**Figure 7.** (a) The distribution profiles of counterion number density, and (b) the distribution profiles of electrostatic potential predicted by **Eq. (S9)** and **Eq. (S6)**, respectively, under different basal spacings and temperatures.

Further, the MD results of swelling pressure isotherms are compared with the classic DLVO calculations. The solid lines ($P_s = P_{PB} + P_{vdW}$) in **Figure 6**(b) are the swelling pressures calculated from the classic DLVO theory, using **Eq. (S11)**. The two components, $P_{PB}$ and $P_{vdW}$ are computed by **Eqns. (S10)** and **(S12)**, and shown by dashed and dotted lines. Firstly, it is found from **Figure 6**(b) that the temperature effect on the swelling pressure predicted by the classic DLVO theory is opposite to that of MD results, that is, the swelling pressure increases with rising temperature. This is due to the thermally enhanced ion correlation effect, which tends to reduce the repulsion ($P_{PB}$) between two charged slabs [12]. The ion correlation effect will also result in an overestimated $P_{PB}$ at larger d-spacings [27], as shown in **Figure 6**(b). This effect is not considered in DLVO theory and therefore the swelling pressure is overestimated. Secondly, **Figure 6**(b) shows that all results from MD simulations are generally enclosed by the curves of $P_{PB}$ and $P_{vdW}$, indicating the possible changing dominance of these interactions with temperature. Thirdly,

**Figure 6**(b) shows that for $d > 3.0$nm, $P_{\text{vdW}}$ is relatively small so $P_s$ is governed by $P_{\text{PB}}$. For shorter distances, $P_{\text{vdW}}$ becomes non-negligible and the shape of the final pressure curves are dominated by $P_{\text{vdW}}$ (the dot lines).

In fact, according to the quantified liquid structure in **Figure S2** and a previous MD simulation study [12], at ambient temperature, the basal spacing $d \approx 2.3$nm separates the clay swelling process of the Na-montmorillonite into two different stages: crystalline swelling and osmotic swelling [44]. In crystalline swelling ($d < 2.3$ nm), the hydration force of interlayer counterions and the negatively charged surfaces drive the swelling process, whereas the osmotic effect dominates in the latter one ($d > 2.3$ nm), as described by the DLVO theory. **Figure 6**(b) indicates that for Na-montmorillonite at large basal spacings ($d > 2.3$nm) and at 298 K, the DLVO theory predictions are generally consistent with the measured repulsive swelling pressures, while the DLVO theory based on the mean-field approximation fails to describe the swelling behaviour at smaller basal spacings ($d < 2.3$nm). These observations are consistent with the previous studies [45], and the discrepancy reflects the increasing importance of the molecular details of the water molecules and counterions, which gives additional forces omitted by DLVO theory and operating at the range of several nanometres. Notably, DLVO theory neglects three types of non-DLVO interparticle interactions [43]. The first is the short-range hydration repulsion induced by surface and counterion hydration. The second is the relatively long-range attraction induced by specific electrostatic interactions between surface charge sites and counterions, counterions and counterions, and surface charge sites and surface charge sites, in overlapping EDLs. The third is induced by the water layering structures, usually oscillating with a period of one water molecule's diameter, i.e., 0.3 nm. **Figure 6**(b) shows that non-DLVO interactions control the swelling pressure for $d < 2.3$ nm at 298 K, and the repulsive pressure from these non-DLVO interactions is weakened with temperature increasing from 298 to 500 K.

Further, the swelling pressure of the Na-MMT at a very high temperature, i.e., 500 K, cannot be reproduced at all by the DLVO theory, where $P_s = P_{PB} + P_{vdW}$. However, the trend within the range $d > 1.7$ nm is well described by the van der Waals component, $P_s = P_{vdW}$. This is because the interlayer counterions dehydrate themselves and tend to adsorb on the negatively charged clay surfaces at 500 K. In such cases, the osmotic pressure, $P_{PB}$, of DLVO interactions for the interlayer molecular body, is significantly decreased. Meanwhile, the non-DLVO interactions, especially the hydration repulsion induced by surface and counterion hydration, tend to be trivial at 500 K and $d > 1.7$ nm due to the deterioration of the layering liquid structure on clay surfaces and dehydrated interlayer counterions. Therefore, $P_s$ can be solely quantified by $P_{vdW}$ within the range $d > 1.7$ nm. For $d < 1.7$ nm, the interlayer counterions enclosed by the implicit walls and the adsorbed water layer directly contacting the clay surface resist being removed and squeezed out by compression. This results in the increase of swelling pressure in the range of $d < 1.7$ nm.

**Extended DLVO model predictions with short-ranged hydration repulsive interaction included:** The hydration force can be further considered and included in the classic DLVO theory to extend its prediction capability for shorter d-spacings. For $d < 2.3$ nm non-DLVO hydration pressure, $P_{\text{hydration}}$, is essential, while the DLVO theory's osmotic pressure, $P_{PB}$, becomes negligible due to highly structured interlayer liquid and consequently decreased entropy and weakened osmotic effect. Therefore, the swelling pressure for $d < 2.3$ nm can be described by $P_s = P_{\text{hydration}} + P_{\text{vdW}}$. By fitting the MD result for the hydration pressure using a monotonic decaying exponential function [46] (see details in **Section 5** in **Supporting Information**), **Figure 6**(c) shows a fair theoretical prediction of such an extended DLVO model for $d < 2.3$ nm.

In addition to the primary monotonic hydration force, a secondary oscillatory structural hydration force can be superimposed to improve the prediction capability of such an extended DLVO model [46]. However, this will require more simulation data to enable an acceptable fitting of oscillation behaviour. This is a topic of ongoing investigation. The fluctuating and stepwise behaviour of the swelling pressure around the curve of $P_s = P_{\text{hydration}} + P_{\text{vdW}}$ can be observed in **Figure 6**(c), especially at shorter separations, reflecting the role of the oscillating hydration force in this regime. The layer-by-layer interfacial liquid structure is destroyed and squeezed out into the bulk liquid when the two surfaces approach each other. This has been discussed in detail in the last section.

Notably, similar results have been previously reported by Yang et al. [27] on Ca-montmorillonite (Ca-MMT) at high temperatures, where the same model and setup have been applied, but with fully rigid clay slabs fixed at certain basal d-spacings without springs. A comparison between the results of this work and [27] is given in **Figure S8** in the **Supporting Information.** It shows that the measured swelling pressures in both works are in the same order of magnitude and that increasing temperature decreases the swelling pressure due to the thermally weakened hydration interaction and enhanced counterion correlation effect. However, the results in [27] show a more pronounced oscillation of swelling pressure, especially at shorter d-spacings. This is because the spring setup used in the present study allows for attaining specific basal d-spacings that correspond to stable interlayer hydration structures (local minima in the free energy curves of clay swelling), while the fixed clay slabs in [27] allow for determining the swelling pressure at certain basal d-spacings with metastable interlayer liquid structure. Therefore, the dependence of the swelling pressure on the basal spacing obtained in this work is smoother than the one obtained in [27].

## 4. Conclusions

Using molecular dynamics simulations, the swelling behaviour of the Na-montmorillonite and the interfacial water under elevated temperatures was studied. It can be concluded that:

1) MD simulations reproduce the experimentally observed reduction of the swelling capacity of clay microstructure at elevated temperatures.

2) The attractive van der Waals force and the repulsive hydration force instead of the repulsive electrostatic (double layer) force dominate the swelling behaviour at high temperatures. The thermally reduced swelling pressure of water-saturated Na-MMT is related to the weakening hydration repulsion as a result of the deteriorating interlayer water layer structure and weakening electric double layer repulsion caused by the shrinking electric double layer structure. At high temperatures, the interlayer counterions are dehydrated and move closer to the charged surfaces to form inner-sphere surface complexes, thus shrinking the electric double layer.

3) The classic DLVO theory captures the swelling behaviour of Na-MMT at lower temperatures and larger separations but fails at short basal spacings (smaller than 2.3 nm) and high temperatures. Without the osmotic contribution from the DLVO model, the summation of the van der Waals interaction in DLVO and an additional non-DLVO hydration interaction is found to fairly predict our MD system's swelling at short basal spacings and high temperatures.

The findings of this study provide a fundamental understanding of the swelling behaviour and the underlying molecular-level mechanisms of the clay microstructure under extreme conditions.

**Author information**

Corresponding Authors:

*E-mail: Weiqiang.Chen@manchester.ac.uk (W. Q. Chen), Majid.Sedighi@manchester.ac.uk (M. Sedighi), Andrey.Jivkov@manchester.ac.uk (A. Jivkov).

Notes: The authors declare no competing financial interest.

**Author contributions**

The manuscript was written through the contributions of all authors. All authors have given approval for the final version of the manuscript.

**Conflicts of interest**

There are no conflicts of interest to declare.

**Acknowledgement**

Chen acknowledges the President Doctoral Scholarship Award (PDS Award 2019) from The University of Manchester, UK. The authors acknowledge the assistance provided by the Research IT team for the use of the Computational Shared Facility at The University of Manchester. Jivkov acknowledges gratefully the financial support from the Engineering and Physical Sciences Research Council (EPSRC), UK, via Grant EP/N026136/1.

**Supporting information**

Additional information including the quantified interlayer liquid structure for the basal spacings of 1.53 nm and 2.2 nm, respectively, the effects of the implicit walls, the classic Derjaguin-Landau-Verwey-Overbeek (DLVO) theory, the determination of Hamaker constant from MD simulation, the determination of hydration interaction, and the effects of spring setup.

**Table of contents image**

For Table of Contents Only

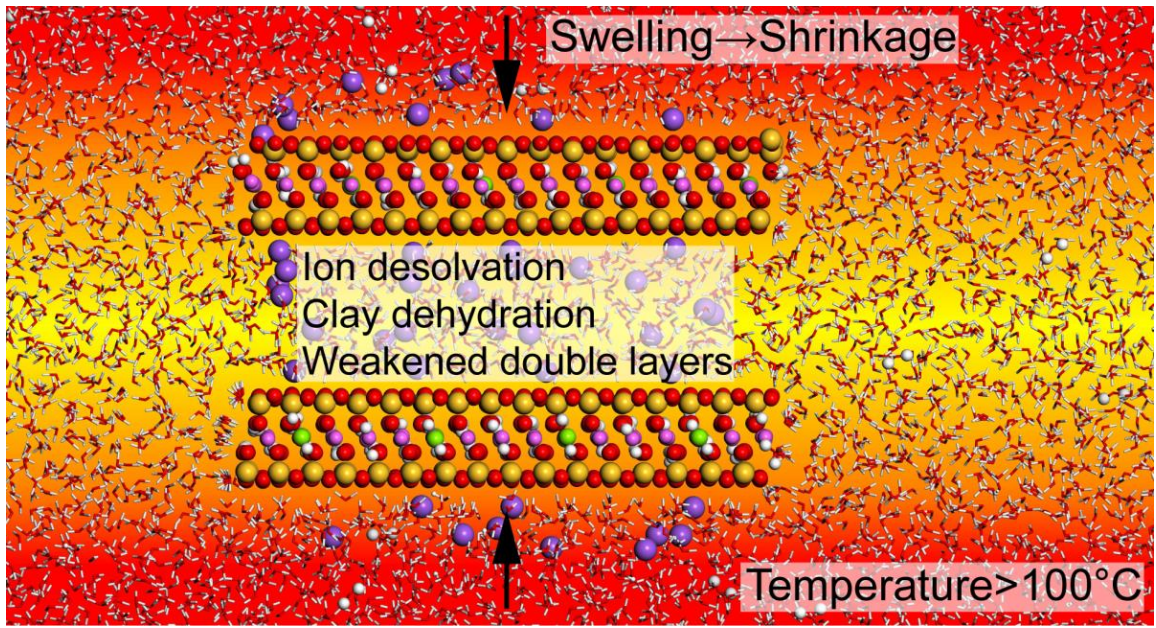

# Supporting Information for

# Elevated temperature effects (T > 100 °C) on the interfacial water and microstructure swelling of Na-montmorillonite

Wei Qiang Chen*, Majid Sedighi*, Florent Curvalle, and Andrey P Jivkov*

School of Engineering, The University of Manchester, Manchester, M13 9PL, United Kingdom

*Corresponding authors:

Weiqiang.Chen@manchester.ac.uk;

Majid.Sedighi@manchester.ac.uk;

Andrey.Jivkov@manchester.ac.uk

## 1. Interlayer liquid structure for d-spacing of 1.53 nm and 2.2 nm

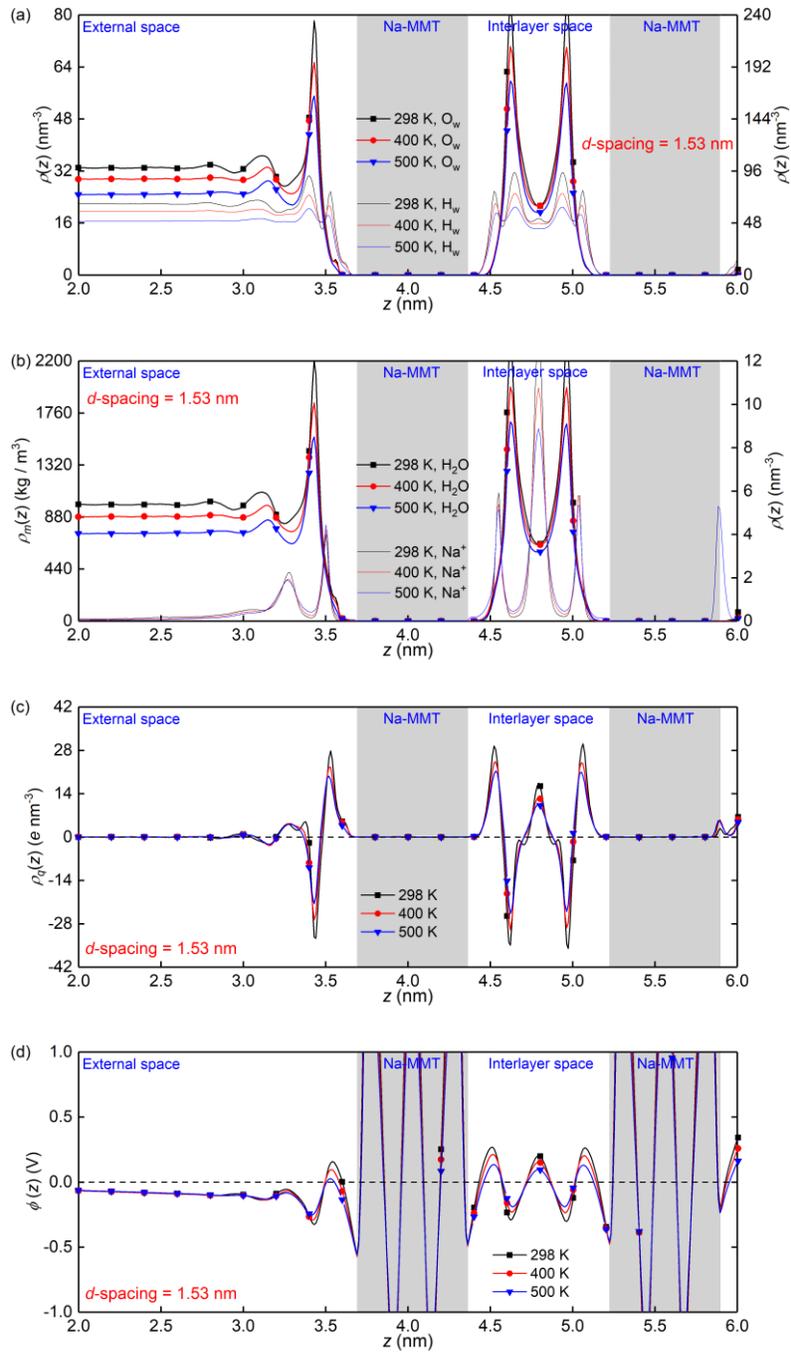

**Figure S1.** (a–b) The number/mass density distributions of water and counterions ($Na^+$) inside and outside the interlayer space of montmorillonite with a d-spacing of 1.53 nm at 50 MPa and different temperatures. (c) The corresponding charge density profiles. (d) The corresponding electrostatic potential of the system along the z direction.

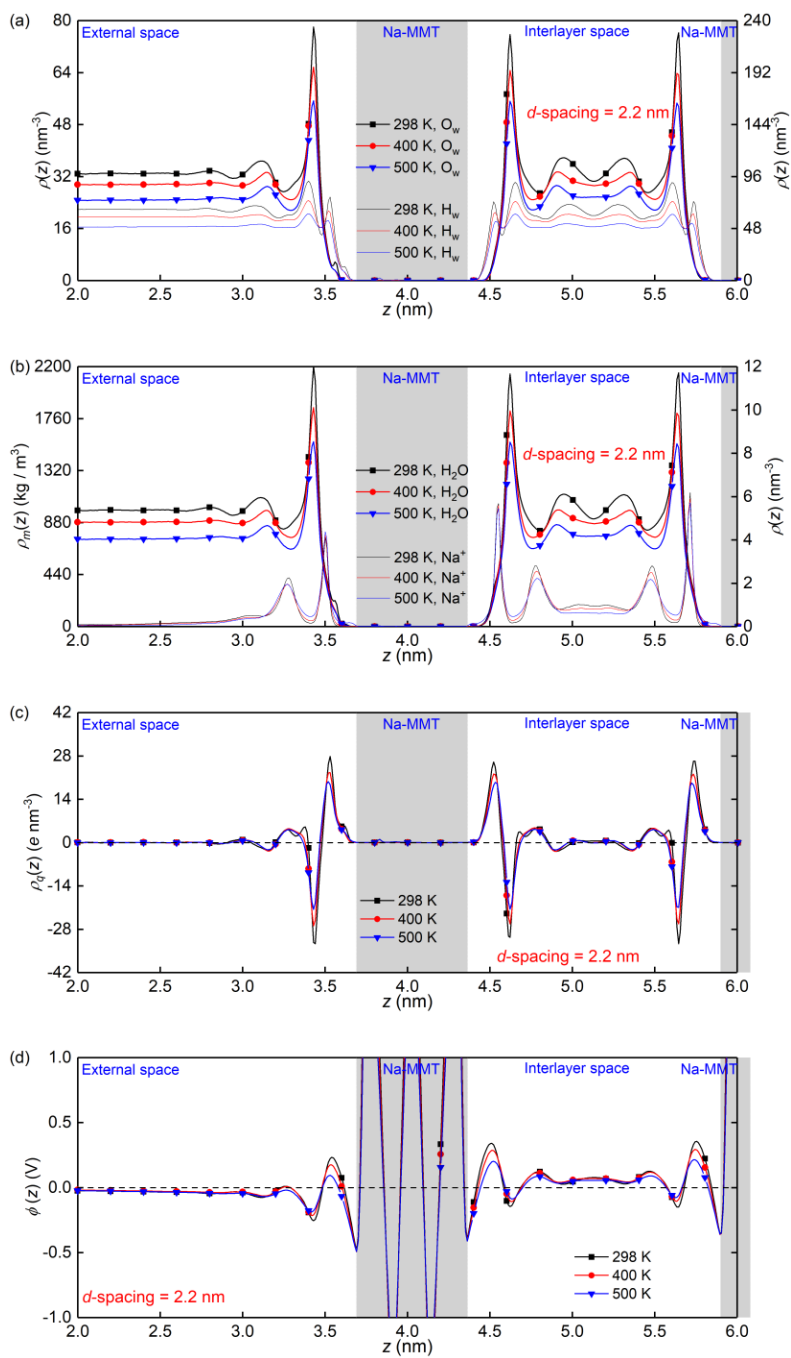

**Figure S2.** (a–b) The number/mass density distributions of water and counterions (Na$^+$) inside and outside the interlayer space of montmorillonite with a d-spacing of 2.2 nm at 50 MPa and different temperatures. (c) The corresponding charge density profiles. (d) The corresponding electrostatic potential of the system along the $z$ direction.

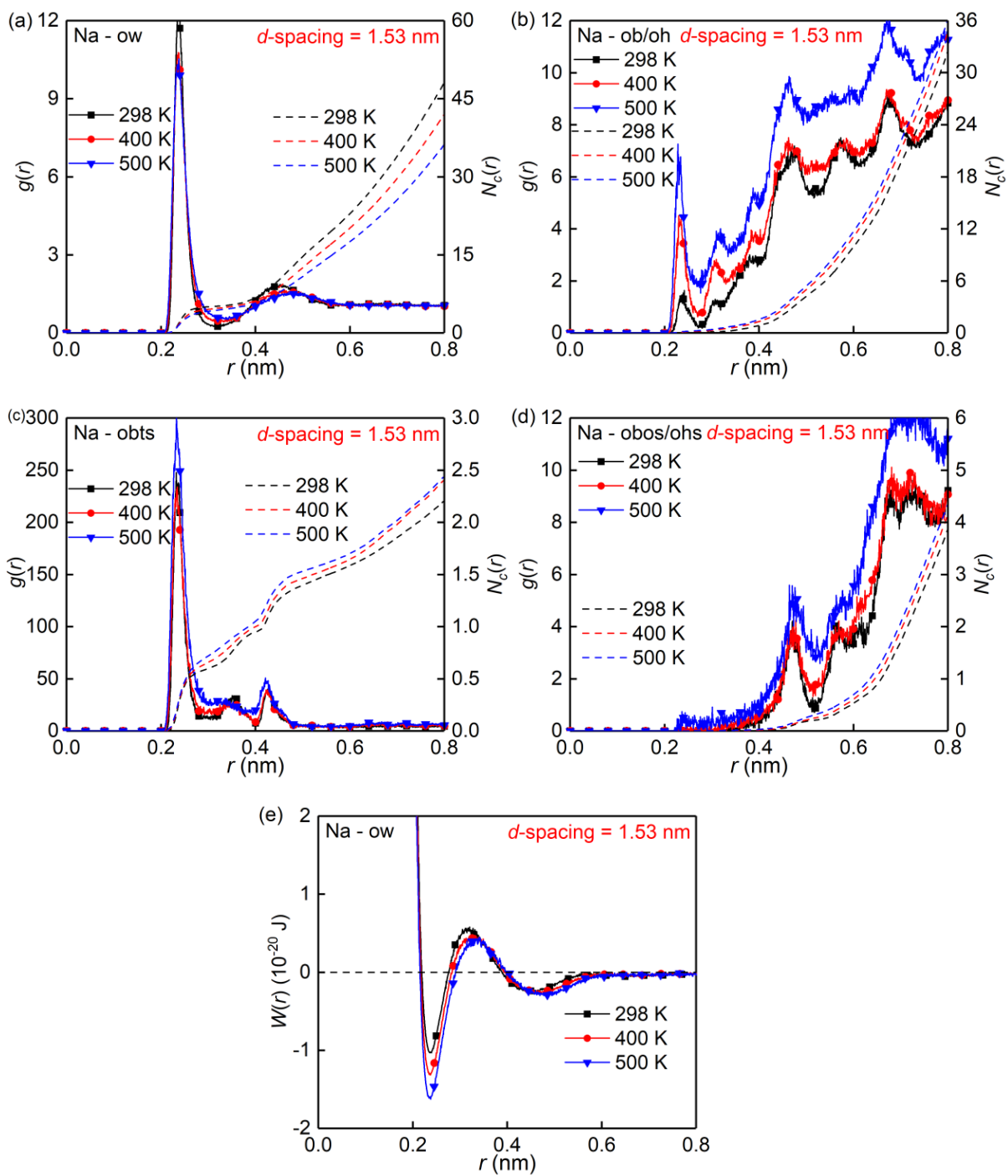

**Figure S3.** (a-d) The radial distribution functions (RDF, solid lines) and coordination numbers (CN, dashed lines) of different pairs with d-spacing of 1.53 nm at 50 MPa and different temperatures. (e) The potential of mean force (PMF) for the Na − ow pair.

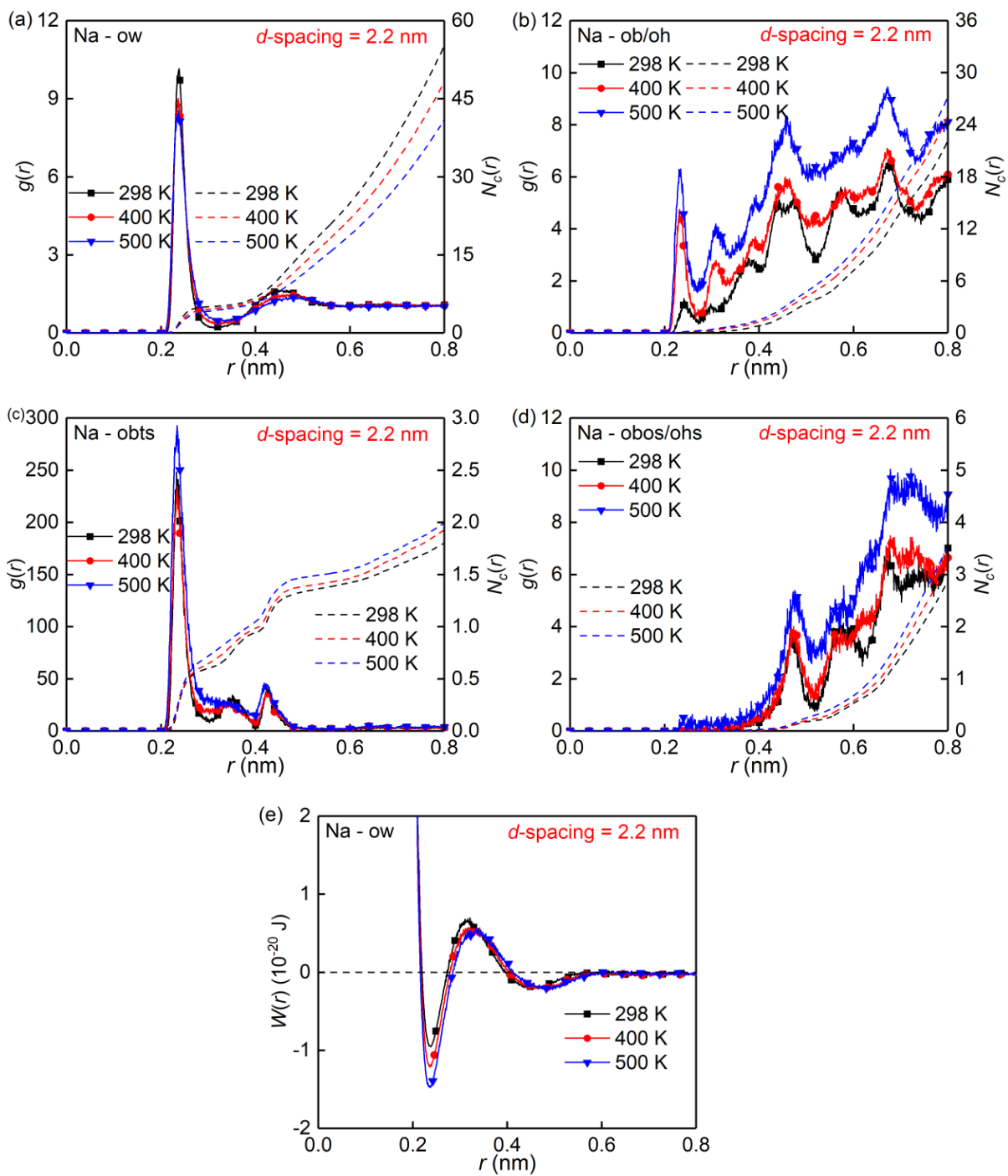

**Figure S4.** (a-d) The radial distribution functions (RDF, solid lines) and coordination numbers (CN, dashed lines) of different pairs with d-spacing of 2.20 nm at 50 MPa and different temperatures. (e) The potential of mean force (PMF) for the Na − ow pair

## 2. Effects of implicit walls

Our MD model adopts the semi-permeable implicit walls to prevent the leakage of the counterions from interlayer spaces to take the osmotic effect into account by maintaining a constant concentration difference between the interlayer space and bulk region. We note that the previous studies have built similar MD models with (e.g., [1]) or without (e.g., [2, 3]) implicit walls. We chose two cases (298 K and 500 K) in our studies to remove the implicit walls and calculate the resulting swelling pressure and equilibrium d-spacings, which are shown in **Figure S5**. The comparisons between cases with and without implicit walls give the generally same results. However, the use of implicit walls gives more stable results (see standard deviations) because the counterions can be effectively confined in the interlayer space and maintain constant local neutral electricity of the Na-MMT.

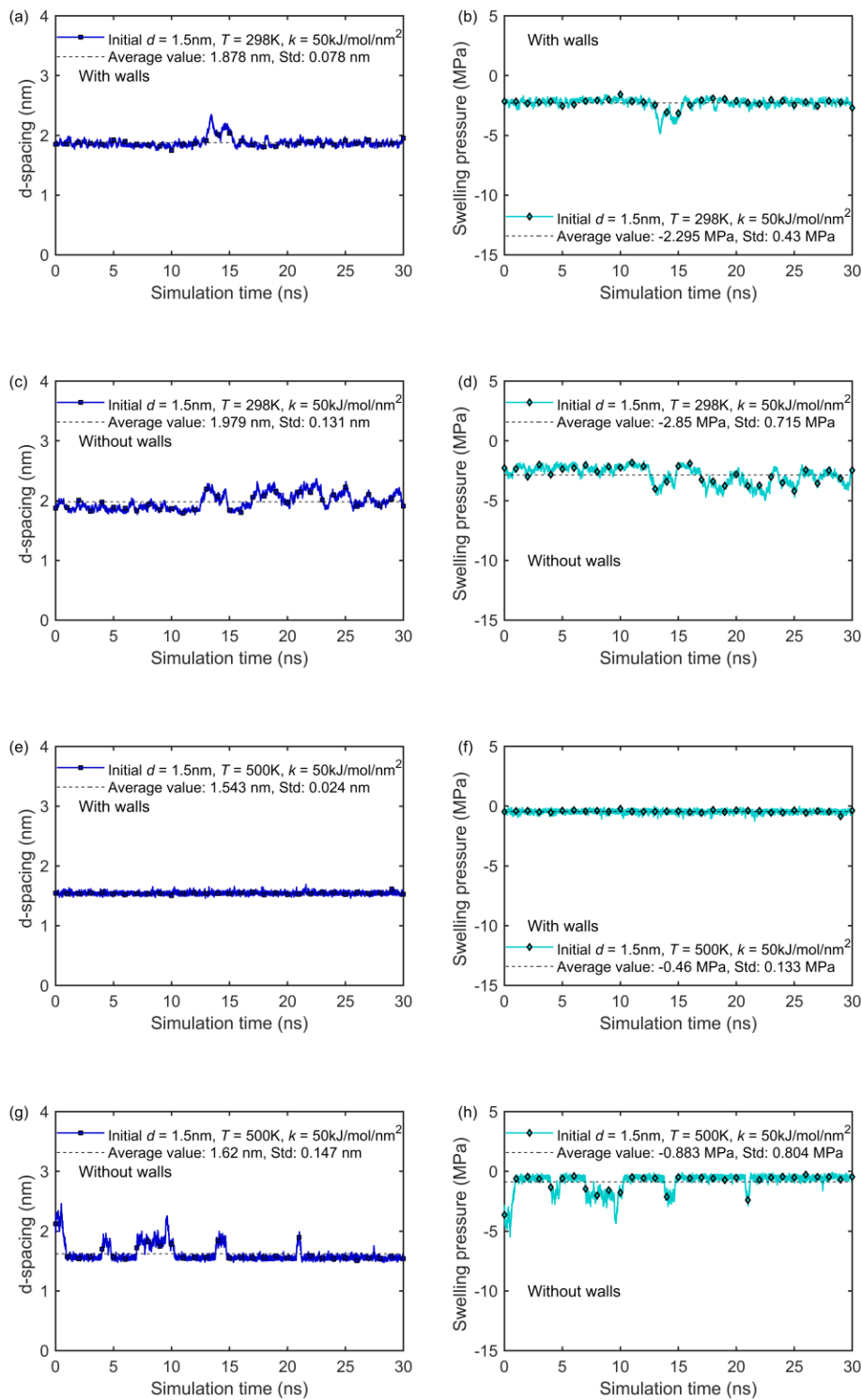

**Figure S5.** The effect of the implicit walls. The left graphs show the results of the equilibrium d-spacings, and the right ones show the equilibrium swelling pressures.

## 3. Derjaguin-Landau-Verwey-Overbeek (DLVO) theory

DLVO theory is the summation of the attractive van der Waals forces and the repulsive electrolytic entropic force [4]. Consider two parallel equally negatively charged planar surfaces separated by a width of $2h$ and forming a slit channel containing only water and one type of counterions, i.e., the solution is mono-ionic. In a coordinate system for which $x = -h$ and $x = h$ are the positions of the two surfaces, the following one-dimensional Poisson-Boltzmann (PB) equation describes the charge distribution [5]:

$$\frac{d^2\phi(x)}{dx^2} = -\frac{Ze}{\varepsilon_r \varepsilon_0} C(0) \exp\left(-\frac{Ze\phi}{k_B T}\right), \tag{S1}$$

where $\phi(x)$ is the electrostatic potential at position $x$, $Z$ is the counterion valency, $e$ is the elementary charge, $\varepsilon_0$ is the vacuum permittivity, $\varepsilon_r$ is the relative dielectric permittivity of water, $C(x)$ is the ionic strength at position $x$, $C(0)$ is the ionic strength at the channel midplane, $k_B$ is Boltzmann's constant, and $T$ is the temperature in Kelvin.

In the middle of the slit channel ($x = 0$), the electric field is zero due to symmetry, i.e.,

$$\left.\frac{d\phi(x)}{dx}\right|_{x=0} = 0. \tag{S2}$$

The boundaries between charged surfaces and interfacial liquid are electroneutral, i.e.,

$$\left.\frac{d\phi(x)}{dx}\right|_{x=h} = -\left.\frac{d\phi(x)}{dx}\right|_{x=-h} = \frac{\sigma}{\varepsilon_r \varepsilon_0}, \tag{S3}$$

where $\sigma$ is the surface charge density. Because the solution contains only counterions, a convenient reference point for the electrostatic potential is:

$$\phi(0) = 0. \tag{S4}$$

Electroneutrality requires that,

$$\int_{-h}^{0} C(x)\mathrm{d}x = \int_{0}^{h} C(x)\mathrm{d}x = \frac{-\sigma}{Ze}. \tag{S5}$$

Eq. (S1) can be solved analytically by combining Eqns. (S2)–(S5) [6], and the solution for $0 \leq x < h$ is (for $-h < x \leq 0$, $\phi(x)$ and $\frac{\mathrm{d}^2\phi(x)}{\mathrm{d}x^2}$ stay the same while $\frac{\mathrm{d}\phi(x)}{\mathrm{d}x}$ changes sign),

$$\begin{cases} \phi(x) = \frac{2k_BT}{Ze} \ln\left\{\cos\left(\frac{sx}{h}\right)\right\} + \phi(0) \\ \frac{\mathrm{d}\phi(x)}{\mathrm{d}x} = -\frac{2k_BTs}{Zeh} \tan\left(\frac{sx}{h}\right) \\ \frac{\mathrm{d}^2\phi(x)}{\mathrm{d}x^2} = -\frac{2k_BTs^2}{Zeh^2} \frac{1}{\cos^2\left(\frac{sx}{h}\right)} \end{cases}, \tag{S6}$$

where the dimensionless parameter $s$ satisfies the relation

$$s \equiv \left(\frac{C(0)(Ze)^2}{2k_BT\varepsilon_r\varepsilon_0}\right)^{1/2} h. \tag{S7}$$

The boundary condition, Eq. (S3), gives

$$s \times \tan(s) = -\frac{Ze\sigma h}{2k_BT\varepsilon_r\varepsilon_0} \equiv K, \tag{S8}$$

where the numerical solution of $s$ is taken within the range of $0 \leq s < \pi/2$ and $K$ is a constant. Therefore, the counterion distribution is

$$C(x) = -\frac{\varepsilon_r\varepsilon_0}{ze} \frac{\mathrm{d}^2\phi(x)}{\mathrm{d}x^2} = 2k_BT\varepsilon_r\varepsilon_0 \frac{s^2}{(zeh)^2} \frac{1}{\cos^2\left(\frac{sx}{h}\right)}. \tag{S9}$$

In addition, because no electrostatic interaction exists between the charged surfaces and the neutral mid-channel, the PB pressure originates only from the osmotic pressure due to counterion concentration in the middle of the slit channel, $C(0)$. This leads to the following expression for the PB pressure [5]:

$$P_{\text{PB}} = k_B T C(0) = \frac{2(k_B T)^2 \varepsilon_r \varepsilon_0}{(Ze)^2} \left(\frac{S}{h}\right)^2. \tag{S10}$$

In the Derjaguin–Landau–Verwey–Overbeek (DLVO) theory [7, 8], the pressure is determined by superimposing the van der Waals attraction ($P_{\text{vdW}}$) onto the osmotic pressure

$$P_s = P_{\text{PB}} + P_{\text{vdW}}, \tag{S11}$$

where $P_{\text{vdW}}$ is calculated by [9]:

$$P_{\text{vdW}} = -\frac{A_H}{6\pi}\left(\frac{1}{8h^3} + \frac{1}{8(h+\delta)^3} - \frac{2}{(2h+\delta)^3}\right), \tag{S12}$$

where $A_H$ is the effective Hamaker constant, which is determined as around $3.8 \times 10^{-20}$ J in **Section** 4 of this supporting information.

Table S1. The parameters for DLVO calculations.

| $T$ (K) | $A_H$ ($10^{-20}$ J) | $\delta$ ($10^{-10}$ m) | $Z$ | $\sigma$ (C/m$^2$) | $\varepsilon_r$ |
|---|---|---|---|---|---|
| 298 |  |  |  |  | 63.58 [1, 11] |
| 400 | 3.8 | 9.6 | 1 | 0.1245 [10] | 40.01 [1, 11] |
| 500 |  |  |  |  | 23.23 [1, 11] |

## 4. Determination of Hamaker constant from MD simulation

The van der Waals forces are generally not computed well by MD simulations that use Lennard-Jones potential with cutoff schemes. Therefore, a Hamaker constant specific to our MD system should be determined instead of using a literature value. By removing the aqueous environment and two pistons in our MD systems, additional simulations were conducted by varying the d-spacing between two clay slabs and extracting the variation of their van der Waals force (pressure) with d-spacing. The Hamaker constant was found to be around $3.8 \times 10^{-20}$ J by fitting MD data with Eq. (S12). Three examples under different temperatures are given in **Figure S6** to show the fitting procedure.

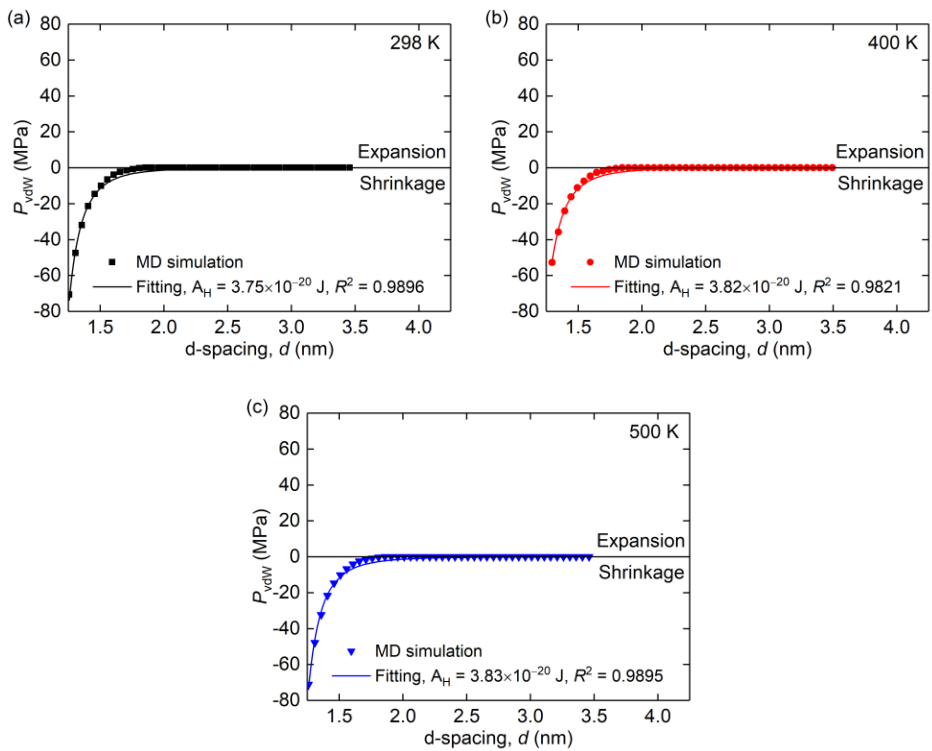

**Figure S6.** Three examples show the fitting procedure to derive the Hamaker constant.

## 5. Determination of hydration interaction

An essential idea regarding the physical mechanisms underlying the hydration force (pressure) is that water molecules form an ordered hydration layer which is strongly bound to the clay surface or counterions. This layer resists removal by compression. An overlap of the water layers near the two approaching clay surfaces or counterions creates a strong short-ranged repulsive force. For this study, the repulsive hydration pressure between two clay surfaces can be described by an empirical exponential decay relation [4]:

$$P_{\text{hydration}} = P_{\text{hydration},0} e^{-(d-\delta)/\lambda}, \tag{S13}$$

where, $P_{\text{hydration},0}$ is the pressure amplitude, $d$ the d-spacing, $\delta = 0.96$ nm the thickness of clay slab, and $\lambda$ the decay length (typically in the range of 0.2–0.4 nm [4]). **Figure S7** shows a good fit of MD result for the hydration pressure by using Eq. (S13), where the hydration pressure is obtained by subtracting the van der Waals pressure calculated by Eq. (S12) from the measured swelling pressure. The result shows a monotonic decline of $\lambda$ with temperature, which is consistent with the observation in the main body of this paper that the adsorbed water layers on clay surfaces become thinner with increasing temperature.

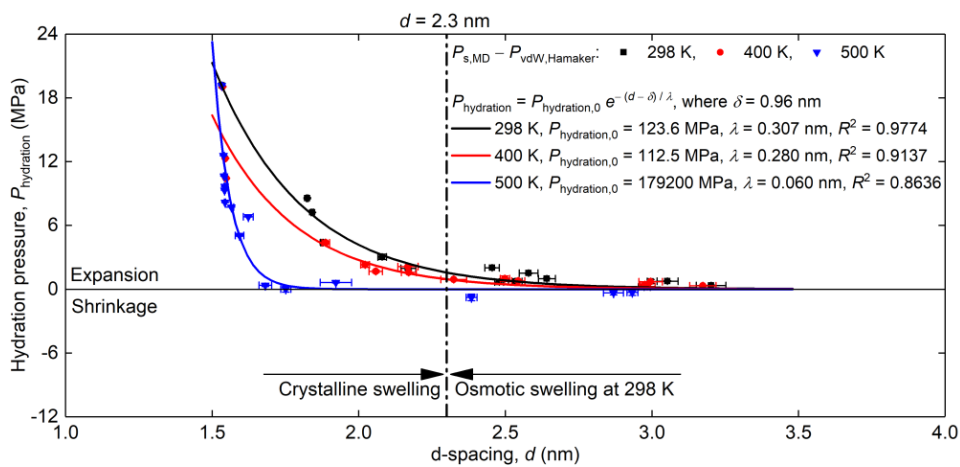

**Figure S7.** The variation of hydration pressure with the d-spacing under different temperatures. The scatters show the MD results for the hydration pressure by deducting the theoretically calculated van der Waals pressure from the measured swelling pressure, while the solid lines show the corresponding curve fitting by Eq. (S13).

## 6. Effects of spring setup

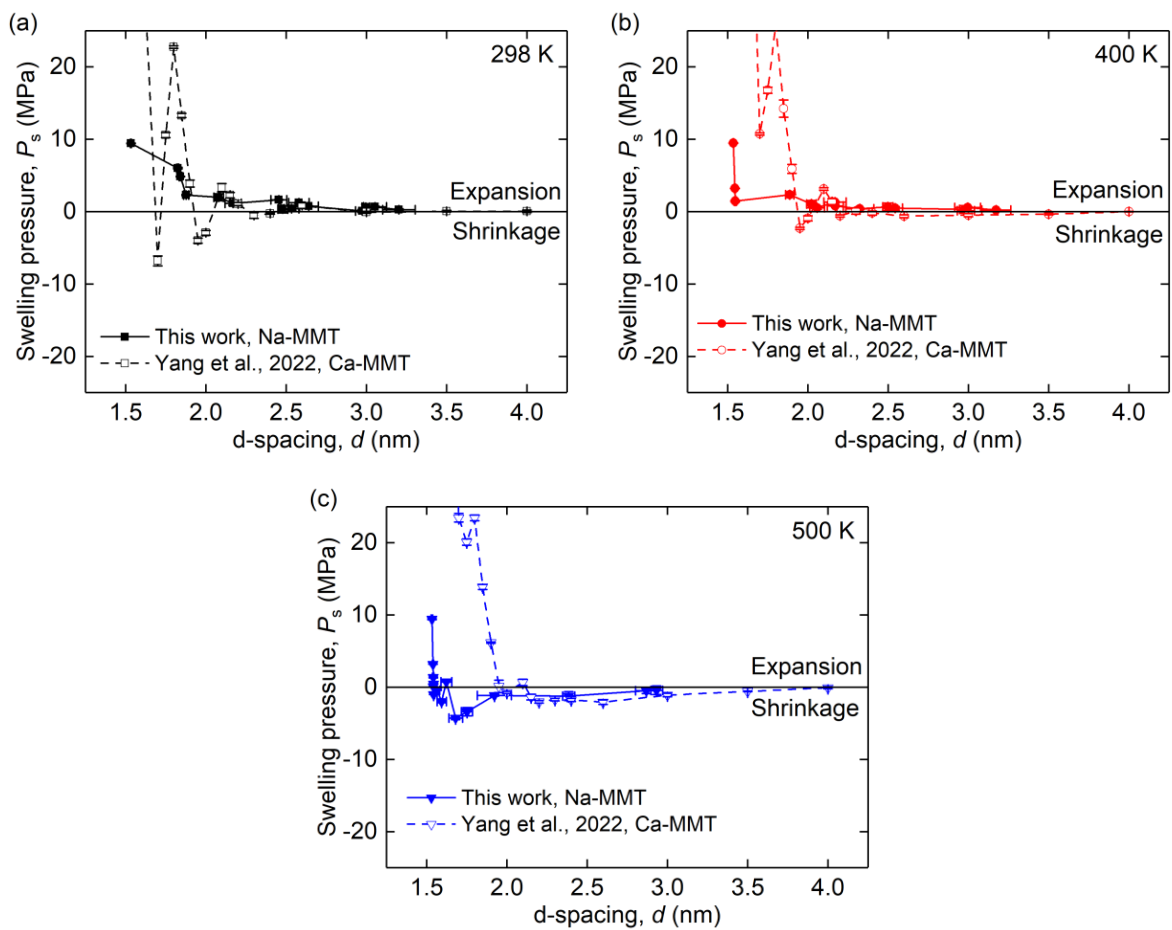

**Figure S8.** Comparisons of swelling behaviour with a previous MD study by Yang et al. [11] on Ca-montmorillonite (Ca-MMT) under different temperatures.